\documentstyle[12pt]{article}

\begin{document}

\title{SUSY GUT Models of $\nu$ Mass and 
$\mu \rightarrow e \gamma$}

\author{{\bf S.M. Barr}\\Bartol Research Institute\\
University of Delaware\\Newark, DE 19716}

\date{}
\maketitle

\begin{abstract}

It is explained why excessive $\mu \rightarrow e \gamma$ can be a problem in
SUSY GUT see-saw models of neutrino mass, and ways that this problem might
be avoided are discussed.

\end{abstract}

\section{SUSY GUT See-saw Models of $\nu$ Mass}

There are three great advantages that SUSY GUT see-saw models have over other
models of neutrino mass and mixing. The first advantage is that neutrino masses
are explained in essentially the same way that the other known fermion
masses are explained, namely the existence of both left- and right-handed
varieties of fermion that have a Dirac coupling to each other through a
Yukawa interaction with a Higgs doublet. GUTs, with the notable
exception of $SU(5)$, generally require that right-handed neutrinos exist. 
Thus they generally {\it predict} that neutrino masses should exist. 
Moreover, GUTs are very well motivated on other grounds. By contrast, in 
many other approaches to explaining neutrino masses they arise from some
postulated new physics that is not strongly motivated on other 
grounds, such as R-parity-violating interactions in SUSY, singly charged scalar
fields (as in the Zee mechanism), or triplet Higgs \cite{bd}.

The second advantage of the SUSY GUT see-saw mechanism is that it explains in
an order of magnitude sense the size of neutrino masses. In the see-saw
formula, the Dirac mass of a neutrino is expected to be related to the
mass of the corresponding up-type of quark. So, if one estimates the mass $m_3$ 
of the third family neutrino, by using the top quark mass for the Dirac
neutrino mass and the GUT scale for the Majorana mass of the right-handed
neutrino, one obtains
\begin{equation}
m_3 \sim  m_{Dirac}^2/M_R \sim (174 \; GeV)^2/(2 \times 10^{16} \; GeV) \sim
1.5 \times 10^{-12} \; GeV.
\end{equation}

\noindent
On the other hand, if neutrino masses are hierarchical, then the
mass $m_3$ can be determined from the mass-squared splitting observed in
atmospheric neutrino oscillations:
\begin{equation}
m_3 \cong \sqrt{m_3^2 - m_2^2} = \sqrt{ \delta m^2_{atm}} \approx 6 
\times 10^{-11} \; GeV.
\end{equation}

\noindent
The agreement is gratifyingly good. By contrast, most other mechanisms for
generating neutrino mass rely upon new physics that involves parameters
that are very poorly if at all constrained by experiment. For example,
we have little idea what the magnitudes of the coefficients of R-parity
violating terms would be, or the couplings of the Zee scalar, or the 
couplings and masses of Higgs triplets. The neutrino masses in such models
can come out to be anywhere within a huge range.

The third advantage of SUSY GUT see-saw models of neutrino mass is that
they can be very predictive, since unified symmetries can precisely relate
the Dirac neutrino mass matrix, the mass matrices of the quarks, and the
mass matrix of the charged leptons. 

Although SUSY GUT see-saw models are extremely well motivated, they face
what at first sight appears to be a serious problem in explaining why
some neutrino mixing angles are large. Since quarks and leptons are unified
in GUTs, there was the expectation for a long time that leptonic mixing 
angles would turn out to be small like the CKM angles. Indeed, that was 
true in GUT models constructed before experimental evidence began to emerge 
in the mid-1990s that the atmospheric angle might actually be large.

More specifically, many models assume hierarchical mass matrices, i.e. the
elements get smaller going up and to the left of diagonal elements.
For instance, consider the classic idea proposed independently by Wilczek-Zee, 
Weinberg, and Fritzsch in 1977 \cite{wwzf}, 
whereby the down quark mass matrix has
the form
\begin{equation}
M_D = \left( \begin{array}{cc} 0 & \epsilon \\ \epsilon & 1 \end{array} \right)
\; m, \;\;\;\; \epsilon \ll 1.
\end{equation}

\noindent
This gives $\tan \theta \cong \epsilon$ and $m_d/m_s \cong \epsilon^2$,
where $\theta$ is the angle required to diagonalize $M_D$.
In fact, it gives exactly $\tan \theta = \sqrt{m_d/m_s}$. If we ignore
the contribution to the Cabibbo angle coming from diagonalizing the
up quark mass matrix $M_U$, then one reproduces the famous empirical
relation $\tan \theta_C \cong \sqrt{m_d/m_s}$.

There are countless models in the literature where the small mass ratios
($m_e \ll m_{\mu} \ll m_{\tau}$, $m_u \ll m_c \ll m_t$, and 
$m_d \ll m_s \ll m_b$) and small CKM angles are explained by assuming that
the mass matrices of the charged leptons ($M_L$), up quarks ($M_U$), and
down quarks ($M_D$) are hierarchical. 

The trouble is that in unification based on $SO(10)$ or larger groups
the Dirac mass matrix of the neutrinos ($M_N$) is related to these 
other mass matrices, and thus might be expected to be hierarchical also. 
However, if $M_N$ is hierarchical, then by the see-saw formula
$M_{\nu} = M_N M_R^{-1} M_N^T$ it would typically be the case that
the light neutrino mass matrix $M_{\nu}$ would be hierarchical too,
leading to small leptonic mixing angles.

For instance, if the Dirac mass matrix of the neutrinos has the hierarchical
form
\begin{equation}
M_N \sim \left( \begin{array}{ccc} 
\epsilon^{\prime 2} & \epsilon \epsilon' & \epsilon' \\
\epsilon \epsilon' & \epsilon^2 & \epsilon \\
\epsilon' & \epsilon & 1 \end{array} \right) \; m,
\end{equation}

\noindent
where $\epsilon$ and $\epsilon'$ are small parameters,
and the Majorana mass matrix $M_R$ has a ``random" form, i.e. all its
elements being of the same order, then the matrix $M_{\nu}$ is easily
seen to have the same hierarchical form as $M_N$, and the
leptonic mixing angles will be small.

There is now strong evidence that two of the three leptonic mixing angles are
large: $|U_{e2}| \equiv \sin \theta_{sol} \approx 0.5$, and $|U_{\mu 3}|
\equiv \sin \theta_{atm} \cong 0.7$. There are 
two main ways to explain these large angles
within the framework of the see-saw mechanism in SUSY GUTs. These have been
called ``Single Right-Handed Neutrino Dominance" or SRND \cite{srnd}, and 
``Lopsided Models" \cite{bb,abb,sy,ilr}. 

\vspace{0.2cm}

\noindent
{\it (A) Single Right-Handed Neutrino Dominance}

\vspace{0.2cm}

\noindent
(i) $\nu_{R2}$ or $\nu_{R1}$ Dominates in See-saw Formula.

This idea is best explained through a simple example. Suppose that the
matrices $M_N$ and $M_R$ have the following forms:
\begin{equation}
M_N \sim \left( \begin{array}{ccc} \eta & \eta & \eta \\ 
\eta & \epsilon & \epsilon \\ \eta & \epsilon & 1 \end{array} \right)m,
\;\;\;\; M_R^{-1} \cong \left( \begin{array}{ccc} 0 & 0 & 0 \\ 0 & 1 & 0 \\ 
0 & 0 & 0 \end{array} \right) \; m_R,
\end{equation}

\noindent
where $\eta \ll \epsilon \ll 1$.
Then it is clear that when these are inserted into the see-saw formula 
the matrix $M_R^{-1}$ approximately projects out the second column of $M_N$,
giving
\begin{equation}
M_{\nu} \sim \left( \begin{array}{c} \eta \\ \epsilon \\ \epsilon \end{array}
\right) \left( \begin{array}{ccc} \eta & \epsilon & \epsilon \end{array}
\right)m^2/m_R \sim 
\left( \begin{array}{ccc} \eta^2 & \eta \epsilon & \eta \epsilon \\
\eta \epsilon & \epsilon^2 & \epsilon^2 \\ \eta \epsilon & \epsilon^2 
& \epsilon^2 \end{array} \right) m^2/m_R.
\end{equation}

\noindent
One sees that as a result of the similar magnitudes of the 22 and 32 elements
of $M_N$, the elements of the 23 block of $M_{\nu}$ are all of the same order,
leading to a large atmospheric angle. Moreover, the fact that $M_R^{-1}$
has nearly the form of a projection matrix of rank one, leads the 23 block
of $M_{\nu}$ to have a nearly ``factorized form", i.e. to be approximately
a rank one matrix. This gives the hierarchy between the neutrino masses
$m_2 \ll m_3$.

\vspace{0.2cm}

\noindent
(i) $\nu_{R3}$ Dominates in See-saw Formula.

Here the idea is that the matrices $M_N$ and $M_R$ have the forms:
\begin{equation}
M_N \sim \left( \begin{array}{ccc} - & - & \epsilon \\ 
- & - & 1 \\ - & - & 1 \end{array} \right) m, \;\;\;\;
M_R^{-1} \cong \left( \begin{array}{ccc} 0 & 0 & 0 \\ 0 & 0 & 0 \\ 
0 & 0 & 1 \end{array} \right) \; m_R,
\end{equation}

\noindent
where the dashes represent much smaller entries. This yields
\begin{equation}
M_{\nu} \sim \left( \begin{array}{ccc} \epsilon^2 & \epsilon & \epsilon \\
\epsilon & 1 & 1 \\ \epsilon & 1 
& 1 \end{array} \right) m^2/m_R.
\end{equation}

\noindent
Here $M_N$ has a large off-diagonal element. This kind of model is similar
in some respects to the lopsided models about to be discussed. Indeed,
in lopsided models, if one goes to a basis where $M_L$ is diagonal, the
matrix $M_N$ acquires the form shown in Eq. (7).

\vspace{0.2cm}

\noindent
{\it (B) Lopsided Models}

\vspace{0.2cm}

The basic insight behind lopsided models is that $SU(5)$ unification
unifies left-handed down quarks with right-handed charged leptons,
and right-handed down quarks with left-handed charged leptons. Consequently,
what are related to the CKM angles by $SU(5)$ unification are the mixing
angles of the {\it right}-handed leptons, {\it not} the observed neutrino 
angles, which of course
are mixings of the left-handed leptons. Thus there is really no discrepancy
in the fact that all of the CKM angles are small and some of the neutrino 
angles are large. 

Suppose that the mass matrix of the charged leptons $M_L$ is highly left-right
asymmetric. In particular, suppose that its 23 element is of the same order
as its 33 element, while its 32 element is smaller by a factor $\epsilon \ll 1$.
Now, $SU(5)$ relates $M_L$ to the transpose of $M_D$. Indeed, in minimal
$SU(5)$ one has $M_L = M_D^T$. This is just due to the fact, already mentioned,
that $SU(5)$ relates down quarks and charged leptons of opposite chirality.
Therefore, in the situation we are considering, if there is an $SU(5)$
symmetry governing the structure of the mass matrices, 
the matrix $M_L$ will have
its 32 element and 33 element being of the same order, while its 23 element
will be a factor of order $\epsilon$ smaller. Notice, then, that the 23 
elements, which are the ones that control the mixing of the second and
third families of left-handed fermions (the 32 elements control the
corresponding right-handed mixing) will be large for the charged leptons 
but small for the down quarks. Thus, $U_{\mu 3}$ could be big
and $V_{cb}$ small, as observed.

A highly predictive $SO(10)$ example of a lopsided model is given in Ref. 5.
The Dirac mass matrices have the following forms:
\begin{equation}
\begin{array}{ll}
M_U = \left( \begin{array}{ccc} \eta & 0 & 0 \\
0 & 0 & -\epsilon/3 \\ 0 & \epsilon/3 & 1 \end{array} \right) m_U, &
M_D = \left( \begin{array}{ccc} 0 & \delta & \delta' \\
\delta & 0 & -\epsilon/3 \\ \delta' & \sigma + \epsilon/3 & 1 
\end{array} \right) m_D, \\ & \\
M_N = \left( \begin{array}{ccc} \eta & 0 & 0 \\
0 & 0 & \epsilon \\ 0 & -\epsilon & 1 \end{array} \right) m_U, &
M_L = \left( \begin{array}{ccc} 0 & \delta & \delta' \\
\delta & 0 & \sigma + \epsilon \\ \delta' & - \epsilon & 1 
\end{array} \right) m_D, \end{array}
\end{equation}

\noindent
where $\eta \ll \delta, |\delta'| \ll \epsilon \ll \sigma \sim 1$. 
It is the large lopsided entry $\sigma$
that leads to large atmospheric neutrino mixing. The large solar mixing
in this model arises from the matrix $M_{\nu}$. In this model there are
six effective Yukawa operators, and seven real parameters ($m_U/m_D$, 
$\sigma$, $\epsilon$, $\delta$, $|\delta'|$, arg$\delta'$, and $\eta$) to
explain 14 measured quantities (8 mass ratios of quarks and charged leptons,
4 CKM parameters, and 2 of the neutrino mixing angles). 

In the model just discussed, the large atmospheric angle comes from the
charged lepton mass matrix, while the large solar angle comes from the neutrino
mass matrix. Another possibility is that both large angles come from the 
charged lepton mass matrix, as happens in ``doubly lopsided models" \cite{dl}. 
In doubly lopsided models the entire third column of the matrix $M_L$ is
large. The large 13 element gives the large solar angle, while the large
23 element gives the large atmospheric angle.

\section{The Problem of $\mu \rightarrow e \gamma$}

It is well known that SUSY GUT models with high scale SUSY breaking can
lead to flavor-changing processes in the lepton sector \cite{fc}. 
It is easy to see why. Consider the superpotential of the MSSM augmented 
to include the right-handed neutrinos:

\begin{equation}
\begin{array}{ccl}
W & = & (Y_U)_{ij} \hat{Q}_i \hat{U}_j \hat{H}_2 + 
(Y_D)_{ij} \hat{Q}_i \hat{D}_j \hat{H}_1 \\
& + & (Y_N)_{ij} \hat{L}_i \hat{N}_j \hat{H}_2 +
(Y_L)_{ij} \hat{L}_i \hat{E}_j \hat{H}_1 \\
& + & \frac{1}{2} (M_R)_{ij} \hat{N}_i \hat{N}_j + \mu \hat{H}_1 \hat{H}_2.
\end{array}
\end{equation}

\noindent
In the basis where $Y_L$ and $M_L$ are diagonal, the matrix $Y_N$ will
in general have off-diagonal elements. Let the slepton masses be degenerate
at a scale $M_*$ that is at or above the unification scale $M_G$:
$m^2_{ij}(M_*) = m^2_0 \delta_{ij}$. The off-diagonal elements in $Y_N$
will induce off-diagonal elements in $m^2$ through the renormalization group
running between $M_*$ and the scale of the right-handed neutrino masses,
which we will call $m_R$. The
induced off-diagonal elements are given in the leading logarithm 
approximation by
\begin{equation}
(\delta m^2)_{ij} \simeq \frac{3 m^2_0 + A^2}{8 \pi^2} \sum_{\ell}
(Y_N)^*_{i \ell} (Y_N)_{j \ell} \ln(M_*/m_R).
\end{equation}

\noindent
The off-diagonal elements $(\delta m^2)_{12}$ and $(\delta m^2)_{23}$ will
lead, when inserted into one-loop diagrams, to the decays $\mu \rightarrow
e \gamma$ and $\tau \rightarrow \mu \gamma$ respectively. These effects 
have been calculated by several groups \cite{fccalc}.
One sees from the previous equation that if
$(Y_N)_{13}$ and $(Y_N)_{23}$ are large at scales above $m_R$, there
will be a significant contribution to $(\delta m^2)_{12}$, i.e. the mixing
of $\tilde{\mu}_L \tilde{e}_L$, and thus to $\mu \rightarrow e \gamma$.

Let us consider the lopsided model shown before. One starts in the original
basis with
\begin{equation}
Y_N \simeq \left( \begin{array}{ccc} \eta & 0 & 0 \\
0 & 0 & \epsilon \\ 0 & -\epsilon & 1 \end{array} \right) h_t, 
\;\;\; M_L = \left( \begin{array}{ccc} 0 & \delta & \delta' \\
\delta & 0 & \sigma + \epsilon \\ \delta' & - \epsilon & 1 
\end{array} \right) m_D,
\end{equation}

\noindent
where $h_t$ is the top quark Yukawa coupling, which is of order one.
The matrix $M_L$ can be diagonalized in stages, the first being to
eliminate the large 23 element $\sigma + \epsilon \sim 1$ that is the
source of the large atmospheric neutrino mixing angle. Rotating in the
23 plane of the left-handed lepton doublets by an angle $\theta_{23} \simeq
\theta_{atm}$ one obtains
\begin{equation}
Y_N \simeq \left( \begin{array}{ccc} \eta & 0 & 0 \\
0 & \epsilon s_{atm} & -s_{atm} \\ 0 & -\epsilon c_{atm} & c_{atm} 
\end{array} \right) h_t, 
\;\;\; M_L = \left( \begin{array}{ccc} 0 & \delta & \delta' \\
\overline{\delta} & \epsilon s_{23} & 0 \\ \overline{\delta'} & -\epsilon c_{23}
& 1/c_{23} 
\end{array} \right) m_D.
\end{equation}

\noindent
We see that there is a large element $(Y_N)_{23}$ directly stemming from
the large atmospheric mixing. The remaining steps in the diagonalization of 
$M_L$ will involve a rotation in the 12 plane of the left-handed lepton
doublets to eliminate the 12 element $\delta$. This rotation is in fact
just analogous to the Cabibbo rotation of the quarks, so we will call the 
rotation angle the ``leptonic Cabibbo angle", and denote it $\theta_C^{\ell}$. 
Just as $\tan \theta_C \simeq \sqrt{m_d/m_s}$, one expects in general that
$\tan \theta_C^{\ell} \sim \sqrt{m_e/m_{\mu}}$. Indeed, in this particular
model $\tan \theta_C^{\ell} \cong \sqrt{m_e/m_{\mu}}$. 
The result of this rotation on the matrix
$Y_N$ is to give:
\begin{equation}
(Y_N)_{23} \simeq - \cos \theta_C^{\ell} \sin \theta_{atm} h_t, \;\;\;
(Y_N)_{13} \simeq - \sin \theta_C^{\ell} \sin \theta_{atm} h_t.
\end{equation}

\noindent
As a result, in lopsided models the amplitude for $\mu \rightarrow e \gamma$ 
is proportional to 
$\frac{1}{2} \sin 2 \theta_C^{\ell} \sin^2 \theta_{atm} h^2_t
\approx \frac{1}{2} \sqrt{m_e/m_{\mu}}$. That is, it is not suppressed by
small Yukawa couplings or small mixing angles. In Ref. 10 the rate for 
$\mu \rightarrow e \gamma$ was calculated in the illustrative model 
that we have been discussing. 
There it was found that for $\tan \beta = 10$, $m_{1/2} = 250$
GeV, $A_0 = m_0$, and $\mu$ positive the branching ratio for this
process varies between $2 \times 10^{-9}$ (for $m_0 = 100$ GeV) and 
$3 \times 10^{-11}$ (for $m_0 = 1$ TeV) \cite{bi}. The experimental limit is 
$1.2 \times 10^{-11}$ \cite{meg}. Calculations that go beyond the leading 
logarithm approximation have been done \cite{petcov}, and these suggest that
the branching ratio for $\mu \rightarrow e \gamma$ computed in Ref. 10
may be underestimated by about an order of magnitude. The lopsided models
thus have a serious issue to confront.

However, t is not only lopsided models that face this issue. Consider 
SRND models
in which the right-handed neutrino of the third family dominates the
see-saw formula. Such models, as we see from Eq.(7), have $(Y_N)_{23}
\simeq \sin \theta_{atm}$ in the original basis. And, since one typically
expects that there will be a leptonic analog of the Cabibbo angle,
the diagonalization of $M_L$ should lead to the relations in Eq. (14).
In other words, the problem arises whether the large atmospheric neutrino
mixing is the result of a large 23 element in $M_L$ or in $M_N$. 

We see that there are four ingredients of the problem of excessive
$\mu \rightarrow e \gamma$. (1) {\it High scale SUSY breaking}. The foregoing
discussion assumed that at some high scale above $m_R$ there are nonzero
slepton masses. (2) {\it SO(10) or a larger group}. This assumption relates
$(Y_N)_{33}$ to $(Y_U)_{33} \simeq h_t \approx 1$. (3) {\it Large
$\theta_{atm}$ from large 23 element of either $M_L$ (lopsided) or
$M_N$ ($\nu_{R3}$-dominated see-saw)}. This assumption leads to the
result that $(Y_N)_{23} \simeq (Y_N)_{33} \simeq 1$ in the basis where
$M_L$ is diagonal. And (4) {\it Significant leptonic Cabibbo mixing}.
This yields $(Y_N)_{13} \simeq \tan \theta_C^{\ell} (Y_N)_{23}$ in the
basis where $M_L$ is diagonal.

\section{Ways Out of the $\mu \rightarrow e \gamma$ Problem}

Corresponding to the four ingredients that go into making the $\mu \rightarrow
e \gamma$ problem serious, there are at least four
approaches to avoiding or alleviating this problem. We will discuss them
in the reverse order.

\vspace{0.2cm}

\noindent
(4) {\it Small leptonic Cabibbo angle}. In several recent papers \cite{bi,bk}, 
models have
been proposed that give acceptable $\mu \rightarrow e \gamma$ by making
$\theta_C^{\ell}$ small. One example is one of the types of models 
analyzed by T. Blazek and S.F. King in Ref. 14. In this SRND type of model  
$Y_N$ has the following form (in the basis where $M_L$ is diagonal):

\begin{equation}
Y_N \sim \left( \begin{array}{ccc} - & \lambda^n & \lambda^m \\
- & \lambda^n & 1 \\ - & \lambda^n & 1 \end{array} \right) f,
\end{equation}

\noindent
where $m\geq 2$, $\lambda \simeq 0.15$, and $f \sim 1$. The matrix $M_R$ is
assumed to be approximately diagonal. This form for $Y_N$ gives acceptably
small $\mu \rightarrow e \gamma$.  The reason is simple. As comparison of
Eqs. (14) and (15) shows,
these models implicitly assume that 
$\tan \theta_C^{\ell} \sim \lambda^m \leq \lambda^2
\simeq 0.02$. This is to be compared to the typical value of order 
$\sqrt{m_e/m_{\mu}} \cong 0.07$ that arises in predictive SUSY GUT models.
This gives, then, at least an order of magnitude suppression of
$\mu \rightarrow e \gamma$.

Another example is provided by a model presented by Bi in Ref. 10. In this 
model, which is of the ``doubly lopsided" type,
the matrix $M_N$ is approximately diagonal, while $M_L$ has the form

\begin{equation}
M_L = \left( \begin{array}{ccc} 0 & \delta & \sigma \\ - \delta & 0 & 
1 - \sigma \\ 0 & \epsilon & 1 \end{array} \right) \; m,
\end{equation}

\noindent
where $\sigma \cong 0.58$, $\epsilon \cong 0.12$, and $\delta \cong 0.00077$.

This model is quite similar in spirit to the lopsided model described
above. Indeed, if one ``rotates away" the large 13 element one gets
a matrix of very similar form to Eq. (12). Going through the diagonalization
procedure, one easily sees that the leptonic
Cabibbo angle comes out to be of order $\delta/\epsilon \sim 0.01$.

The trouble is that in a GUT model the matrix $M_L$ is intimately related to
the matrix $M_D$. Thus, if this model were to arise from a GUT one would
have to explain why $\sin \theta_C \cong 0.21$, while $\sin \theta_C^{\ell}
\sim 0.01$. Unfortunately, the model as presented in Ref. 10 gives only the 
leptonic matrices, and so is not a complete unified model of quark and
lepton masses and mixings.

To see that it is difficult to make the leptonic Cabibbo angle small in
a predictive SUSY GUT model of quark and lepton masses, let us see what happens
if we attempt to do this by modifying the model of Ref. 5 described earlier.
To allow the leptonic Cabibbo angle to be small let us add some
more parameters. In particular, take the matrices $M_L$ and $M_D$ to have
the forms:
\begin{equation}
M_L = \left( \begin{array}{ccc}
0 & \delta - \overline{\delta} & \delta' - \overline{\delta}' \\
\delta + \overline{\delta} & 0 & \sigma + \epsilon \\
\delta' + \overline{\delta}' & - \epsilon & 1 \end{array} \right) \; m_D,
\;\; 
M_D = \left( \begin{array}{ccc}
0 & \delta + \frac{1}{3}\overline{\delta} & \delta' + \frac{1}{3} 
\overline{\delta}' \\
\delta - \frac{1}{3} \overline{\delta} & 0 & - \frac{1}{3} \epsilon \\
\delta' - \frac{1}{3} \overline{\delta}' & \sigma + 
\frac{1}{3} \epsilon & 1 \end{array} 
\right) \; m_D.
\end{equation}

\noindent
Here the new parameters $\overline{\delta}$ and $\overline{\delta}'$
(both complex) have been added, so that the model has 11 instead of 7
real parameters. These forms can be obtained from fairly simple effective
operators. 

It is clear that the leptonic Cabibbo angle will be suppressed if the
12 element of $M_L$ is sufficiently small. This can be achieved here through a
cancellation between $\delta$ and $\overline{\delta}$. 
Let us define a suppression parameter $\alpha$ by
$\theta_C^{\ell} = \alpha \sqrt{m_e/m_{\mu}}$. Thus, in the model of Ref. 5
one has that $\overline{\delta} = \overline{\delta}' = 0$ and $\alpha = 1$.

Interestingly, one finds that although we have added several parameters, the
model is almost as predictive as before. In fact, it gives all the same 
predictions, essentially, except that the prediction for $V_{ub}$ is altered. 
One finds
\begin{equation}
\frac{V_{ub}}{V_{cb} V_{us}/V_{cs}} \cong \frac{1}{\sigma^2} \left[
-1 + \sqrt{\sigma^2 + 1} ( 2 s_1 - s_2 \alpha) \right],
\end{equation}

\noindent
where $s_1$ and $s_2$ are phases. The model of Ref. 5 corresponds to the case 
$\alpha = 1$, $s_1 = s_2 \equiv e^{i \theta}$, and $\sigma \cong \sqrt{3}$.,
giving $V_{ub}/(V_{cb} V_{us}/V_{cs}) 
\cong -\frac{1}{3} + \frac{2}{3} e^{i \theta}$,
which gives an excellent fit to the data. (It gives a circle in the
complex plane of radius $\frac{2}{3}$ centered at the point $(-\frac{1}{3}, 0)$.
This passes through the experimentally allowed region.)
However, if one tries to suppress the leptonic Cabibbo angle, that means
that $\alpha \ll 1$, giving $V_{ub}/(V_{cb} V_{us}/V_{cs}) \cong -\frac{1}{3} 
+ \frac{4}{3} e^{i \theta}$, which is inconsistent with data by a large
margin, since experimentally $|V_{ub}/(V_{cb} V_{us}/V_{cs})| \simeq 0.4$,
while the prediction obviously implies that
$|V_{ub}/(V_{cb} V_{us}/V_{cs})| \geq 1.0$.

Another way to modify the model of Ref. 5 to allow small leptonic Cabibbo angle
is by changing the entries for the first family as follows:
\begin{equation}
M_L = \left( \begin{array}{ccc}
0 & \delta - \eta & 0 \\
\delta' + \eta & 0 & \sigma + \epsilon \\
0 & - \epsilon & 1 \end{array} \right) \; m_D,
\;\; 
M_D = \left( \begin{array}{ccc}
0 & \delta' + \frac{1}{3}\eta & 0 \\
\delta - \frac{1}{3} \eta & 0 & - \frac{1}{3} \epsilon \\
0 & \sigma + \frac{1}{3}\epsilon & 1 \end{array} 
\right) \; m_D.
\end{equation}

\noindent
Here again one sees that the possibility exists of a suppression of the
leptonic Cabibbo angle through a cancellation between $\delta$ and $\eta$.
The forms in Eq. (19) give a reasonable fit to the masses of the first 
family fermions, and to the Cabibbo angle, but yields the prediction
\begin{equation}
\frac{V_{ub}}{V_{cb} V_{us}/V_{cs}} \cong \frac{1}{\sqrt{1+ \sigma^2}} \left[
2 - \alpha s_1 \right] s_2,
\end{equation}

\noindent
where $\alpha$ is defined as before, and $s_1$ and $s_2$ are 
phases. For $\alpha \ll 1$ (i.e. small leptonic Cabibbo angle) one has 
then that $V_{ub}/(V_{cb} V_{us}/V_{cs})$
is a complex number of unit magnitude, a result completely inconsistent
with experiment.

There are other ways to modify the model of Ref. 5 to make it
more flexible and able to accomodate a small leptonic Cabibbo angle, besides
the two examples just shown.
However, if the models remain predictive, they tend to give a bad fit
to some quantity, usually $V_{ub}$. Moreover, the suppression of the leptonic
Cabibbo angle is achieved generally by an artificial cancellation,
as in these two examples. This does not prove that
predictive SUSY GUT models of quark and lepton masses with small leptonic
Cabibbo angle cannot be constructed, but it suggests that constructing
such models is not a trivial matter.

\vspace{0.2cm}

\noindent
(3) {\it $\theta_{atm}$ 
does not arise from large 23 elements of $M_L$ or $M_N$}.
One can imagine a number of ways that a large atmospheric angle could
arise without there being a large 23 element in either $M_L$ or $M_N$.
One way is through SRND with the first or second family right-handed neutrino
dominating as in Eq. (5). The main question is whether this idea can be 
implemented in predictive SUSY GUT models. There may be some difficulty 
in doing this, since the form of $M_N$ shown in Eq. (5) may clash with
the form of $M_U$ needed to obtain $m_c/m_t \ll V_{cb}$. (Having a form for 
$M_U$ that is similar to the form for $M_N$ given in Eq. (5) would give
$m_c/m_t \sim \epsilon$ and $V_{cb} \sim \epsilon$.)

A second possibility is that a large atmospheric angle might arise from a
non-see-saw mechanism. That is it could be that $M_{\nu} =
- M_N M_R^{-1} M_N^T + M_{\nu}^{non-see-saw}$. Then $M_N$ and $M_L$ could both
have small 23 elements. However, such a scenario sacrifices most of the 
advantages of SUSY GUT see-saw models enumerated in the first section of this
paper.
One would have to invoke new ad hoc physics to generate 
$M_{\nu}^{non-see-saw}$. One would have to explain why the non-see-saw 
mechanism happened to give neutrino masses that were of the same magnitude
(roughly speaking) as the see-saw mechanism. And one would give up any
predictivity for the neutrinos' masses and mixings.

A third possibility is that in fact both $M_L$ and $M_N$ have large 23
elements, but the flavor-changing effects are suppressed by a
cancellation between them. An example of this is the model of
Babu, Pati, and Wilczek \cite{bpw}. In that model, $(Y_N)_{23}/(Y_N)_{33} 
= \sigma - \epsilon$ and $(Y_L)_{23}/(Y_L)_{33} 
= \eta - \epsilon$, where it happens that $\sigma - \epsilon = - 0.395$
and $\eta- \epsilon = -0.436$. Thus, in the basis where $M_L$ ($\propto
Y_L$) is diagonal, the matrix $Y_N$ has a small 23 element by cancellation.
This cancellation does not occur for the atmospheric angle, however, 
since the form of $M_R$ is such that $M_{\nu}$ has a small 23 element.
It should be noted that the same cancellation between $\sigma$ and $\eta$
happens for the quark sector and makes $V_{cb}$ small. One drawback to
this model is that the cancellation is a numerical accident, though perhaps
another version of the model could be found where this accident is explained.

\vspace{0.2cm}

\noindent
(2) {\it An SU(5) or SU(5)-like model}. 
The group $SO(10)$ relates $Y_N$ to $Y_U$ and tends to predict that
$(Y_N)_{33} \sim h_t$, but $SU(5)$ leaves $(Y_N)_{33}$ free to be
much smaller than $h_t$. Unfortunately, $SU(5)$ is much less predictive
for quark and lepton masses than $SO(10)$ or larger groups. 

An interesting possibility is the following. Suppose that $SO(10)$ breaks 
to $SU(5)$ at some scale $M_{10} \sim M_*$ (the scale at which slepton 
masses are degenerate). In the $SO(10)$ theory above $M_{10}$ let the
Dirac Yukawa coupling matrix of the neutrinos be $Y^{(0)}_N$, which
couples the $\nu$ and $N^c$ in the three families ($16_i$) to each other.
Let there also be $SO(10)$-singlet 
fermions $N_i$ in each family which get masses
from the terms $M_{ij} N_i N_j + Y_{ij} N_i^c N_j 
\langle \Sigma(\overline{16}) \rangle$, where both $M$ and $\langle \Sigma
\rangle$ are of order $M_{10}$. Then below $M_{10}$ the remaining 
$SU(5)$-singlet fermions (i.e. ``right-handed neutrinos") are linear
combinations of $N^c_i$ and $N_i$. Thus, the effective Dirac Yukawa
coupling matrix of the neutrinos below $M_{10}$ is $Y_N \cong
Y^{(0)}_N [(Y \langle \Sigma \rangle)^{-1} M] \equiv Y^{(0)}_N \lambda$.
The parameters can easily be such that the matrix $\lambda$ has small
elements. Then the elements of $Y_N$, and in particular the 33 element
can be small. This would suppress flavor changing effects due to running
between $M_{10}$ and $m_R$, as one sees from Eq. (11).

However, above the scale $M_{10}$ there are large Dirac Yukawa couplings
for the neutrinos, since one expects that $(Y^{(0)}_N)_{33} \sim h_t$ by
$SO(10)$ symmetry. Thus, if $M_* \geq M_{10}$, one will still get contributions
to flavor-changing slepton masses, but with $\ln (M_*/m_R)$ in Eq. (11) 
replaced by $\ln (M_*/M_{10})$, which could easily be of order one.
In this way, one suppresses the rate for $\mu \rightarrow e \gamma$ by
one or two orders of magnitude.

\vspace{0.2cm}

\noindent
(1) {\it SUSY breaking occurs at a low scale.} The whole discussion of the
flavor changing problem above is predicated upon the assumption that
there is high-scale SUSY breaking, so that there are slepton masses at a
scale $M_*$ that is large compared to the right-handed neutrino mass scale
$m_R$. In schemes such as gauge-mediated SUSY breaking, there would not be
a problem with $\mu \rightarrow e \gamma$ such as we have been analyzing.

In conclusion, it would seem that most ways to suppress 
$\mu \rightarrow e \gamma$ in the context of high-scale SUSY breaking only
do so to a rather limited extent. If $SO(10)$ is broken to $SU(5)$ at a
scale near that at which slepton masses are degenerate, it gives a
suppression of only an order of magnitude or two in the rate. The leptonic 
Cabibbo angle is difficult to suppress. 
The SRND models in which the right-handed neutrino of the first or second
family dominates could give a strong suppression, but it is doubtful whether
realistic and predictive schemes based on this idea can be constructed; 
so far they have not been. 

By far the cleanest way to get rid of the problem is by some low-scale
SUSY breaking. Perhaps this is also the most plausible possibility.
There is, even if we ignore large neutrino mixings and GUTs, a well-known
problem with flavor changing processes in models with high-scale SUSY
breaking. None of the proposed solutions 
to this ``SUSY Flavor Problem" have so far been very convincing.
One can regard the problem of $\mu \rightarrow e \gamma$ in SUSY GUT
see-saw models as simply a particularly acute form of a more general
problem with high-scale SUSY breaking. It is reasonable to suppose that
whatever mechanism resolves the other flavor-changing problems of SUSY may
also resolve this one.

\end{document}